\documentclass[twocolumn,showpacs,showkeys,prd,aps,amsmath,amssymb,10pt]{revtex4-1}
\usepackage{graphicx}
\usepackage{color}
\usepackage{multirow}
\usepackage{bm}% bold math
\usepackage{mathtools}
\usepackage{subfig}
\begin{document}
\title{Shell model studies of competing mechanisms to the neutrinoless double-beta decay in $^{124}$Sn, $^{130}$Te, and $^{136}$Xe}
\author{Andrei Neacsu}
\email{neacs1a@cmich.edu}
\author{Mihai Horoi}
\email{mihai.horoi@cmich.edu}
\affiliation{Department of Physics, Central Michigan University, Mount Pleasant, Michigan 48859, USA}
\date{\today}
\begin{abstract}
Neutrinoless double-beta decay is a predicted beyond Standard Model process that could clarify some of the not yet known neutrino properties, such as the mass scale, the mass hierarchy, and its nature as a 
Dirac or Majorana fermion. Should this transition be observed, there are still challenges in understanding the underlying contributing mechanisms.
We perform a detailed shell model investigation of several beyond Standard Model mechanisms that consider the existence of right-handed currents.  Our analysis presents different venues that can be used to identify the dominant mechanisms for nuclei of experimental interest in the mass
A$\sim$130 region ($^{124}$Sn, $^{130}$Te, and $^{136}$Xe).
It requires an accurate knowledge of nine nuclear matrix elements that we calculate, in addition to the associated energy dependent phase-space factors.
\end{abstract}
\maketitle
%\tableofcontents

\renewcommand*\arraystretch{1.3}

\section{Introduction}
Should the neutrinoless double-beta decay ($0\nu\beta\beta$) be experimentally observed, the lepton number conservation is violated by two units and back-box theorems 
\cite{SchechterValle1982,Nieves1984,Takasugi1984,Hirsch2006} predict the neutrino to be a Majorana particle. 
In addition to the nature of the neutrino (whether is a Dirac or a Majorana fermion) there are other unknown properties of the neutrino that could be investigated 
via $0\nu\beta\beta$, such as the mass scale, the absolute mass, or the underlying neutrino mass mechanism. There are several beyond Standard Model mechanisms that could compete and contribute
to this process \cite{Vergados2012,Horoi2013}. Reliable calculations of the nuclear matrix elements (NME) are necessary to perform an appropriate analysis that could help evaluate the 
contributions of each mechanism. 

The most commonly investigated neutrinoless $0\nu\beta\beta$ mechanism is the so called ``mass mechanism'' involving the exchange of light left-handed neutrinos, 
for which the (NME) were calculated using many nuclear structure methods. 
Calculations that consider the contributions of heavy, mostly sterile, right-handed neutrinos have become recently available, while left-handed heavy neutrinos have been shown to have a 
negligible effect \cite{Mitra2012,Blennow2010} and their contribution is generally dismissed. 
A comparison of the recent mass mechanism results obtained with the most common methods can be seen in Fig. 6 of Ref. \cite{NeacsuHoroi2016}, where one can notice 
the differences that still exist among these nuclear structure methods. Fig. 7 of Ref. \cite{NeacsuHoroi2016} shows the heavy neutrino results for several nuclear structure methods, 
and the differences are even larger than in the light neutrino case because of the uncertainties related to the short-range correlation effects (SRC). There are efforts to reduce these
uncertainties by the development of an effective transition operator that treats the SRC consistently \cite{HoltEngel2013}.

Because shell model calculations were successful in predicting two-neutrino double-beta decay 
half-lives \cite{Retamosa1995} before experimental measurements, and as shell model calculations of different groups largely agree with each other without the need to adjust model parameters, 
we calculate our nuclear matrix elements using shell model techniques and Hamiltonians that reasonably describe the experimental spectroscopic observables.

Experiments such as SuperNEMO \cite{snemo2010,nemo32014} could track the outgoing electrons and help distinguish between the mass mechanism ($\nu$) and the 
so called $\lambda$, and $\eta$  mechanisms \cite{Doi1985,Barry2013}. 
This would also provide complementary data at low energies for tesing the existence of right-handed contributions predicted by left-right symmetric models 
\cite{PatiSalam1974,MohapatraPati1975,Senjanovic1975,KeungSenjanovic1983,Barry2013}, currently investigated at high energies in colliders and accelerators such as LHC \cite{CMS2014}.
To distinguish the possible contribution of the heavy right-handed neutrino using shell model nuclear matrix elements, measurements of lifetimes for at least two different isotopes are 
necessary, ideally that of an A$\sim$80 isotope and another lifetime of an A$\sim$130 isotope, as discussed in Section V of Ref. \cite{HoroiNeacsu2016prd}.
It is expected that if the neutrinoless double-beta decay is confirmed in any of the experiments, more resources and upgrades could be dedicated 
to boost the statistics and to reveal more information on the neutrino properties.

Following our recent study for $^{82}$Se in Ref. \cite{HoroiNeacsu2016prd}, which is the baseline isotope of SuperNEMO, 
we extend our analysis of the $\lambda$ and $\eta$  mechanisms to other nuclei of immediate experimental interest: $^{124}$Sn, $^{130}$Te, and $^{136}$Xe. These isotopes are under investigation by the TIN.TIN \cite{Vandana2014} ($^{124}$Sn), CUORE \cite{Cuore2015,Cuore2016}, SNO+ \cite{SNO2016} ($^{130}$Te), NEXT \cite{Next2014}, EXO \cite{Exo2014}, and KamlandZEN \cite{Kamlandzen2016} ($^{136}$Xe) experiments.
For the mass region A$\sim$130 we perform calculations in the $jj55$ model space consisting of $0g_{7/2}, 1d_{5/2}, 1d_{3/2}, 2s_{1/2}$ and $0h_{11/2}$ valence orbitals using 
the SVD shell model Hamiltonian \cite{Chong2012} that was fine-tuned with experimental data from Sn isotopes. Our tests of this Hamiltonian include energy levels, 
$B(E2)\uparrow$ transitions, occupation probabilities, Gamow-Teller strengths, and NME decompositions for configurations of protons/neutrons pairs coupled to some
spin (I) and some parity (positive or negative), called $I$-pair decompositions. These tests and validations of the SVD Hamiltonian can be found in Ref. \cite{NeacsuHoroi2016} for $^{124}$Sn and in Ref. \cite{NeacsuHoroi2015} for 
$^{130}$Te and $^{136}$Xe. Calculations of NME in larger model spaces (e.g. the $jj77$ model space that includes the $0g_{9/2}$ and $0h_{9/2}$ orbitals missing in the $jj55$ models space) were 
successfully performed for $^{136}$Xe \cite{HoroiBrown2013}, but for $^{124}$Sn and $^{130}$Te are much more difficult and would require special truncations. 

In this work, assuming the detection of several tens of $0\nu\beta\beta$ decay events, we present a possibility to identify right-handed contributions from the $\lambda$ and $\eta$ 
mechanisms by analysing the two-electron angular and energy distributions that could be measured.

We organize this paper as follows: 
Section II shows a brief description of the neutrinoless double-beta decay formalism considering a low-energy Hamiltonian that takes into account contributions from right-handed currents.
Section III presents an analysis of the half-lives and of the two-electron angular and energy distributions results for $^{124}$Sn, $^{130}$Te, and $^{136}$Xe.
Finally, we dedicate Section IV to conclusions.

\section{Brief formalism of $0\nu\beta\beta$}\label{formalism}

The existence of right-handed currents and their contributions to the neutrinoless double-beta decay rate has been considered for a long time \cite{Doi1983,Doi1985}, but most frequently calculations
considered only the light left-handed neutrino-exchange mechanism (commonly referred to as ``the mass mechanism''). 
One model that considers the right-handed currents contributions and that includes heavy particles that are not part of the Standard Model is the left-right symmetric model \cite{MohapatraPati1975,Senjanovic1975}.
Within the framework of the left-right symmetric model one can the neutrinoless double-beta decay half-life  expression as
\begin{eqnarray}
\label{flifetime}
\nonumber \left[ T^{0\nu}_{1/2} \right] ^{-1}  =  G^{0\nu}_{01} g^4_A & \mid & M^{0\nu}\eta_{\nu} +M^{0N}\left(\eta^L_{N_R}+\eta^R_{N_R}\right) \\
 & & +\eta_{\lambda}X_{\lambda}+\eta_{\eta}X_{\eta}+\cdots \mid ^2  ,
\end{eqnarray}
where $\eta_{\nu}$, $\eta^L_{N_R}$, $\eta^R_{N_R}$, $\eta_{\lambda}$, and $\eta_{\eta}$ are neutrino physics parameters defined in Ref. \cite{Barry2013} (see also Appendix A of Ref. \cite{HoroiNeacsu2016prd}), 
$M^{0\nu}$  and $M^{0N}$ are the light and heavy neutrino-exchange nuclear matrix elements \cite{Horoi2013,Vergados2012}, and $X_{\lambda}$ and $X_{\eta}$ are combinations of NME and phase space factors that are calculated in this paper.
$G_{01}^{0\nu}$ is a phase space factor \cite{SuhonenCivitarese1998} that one can calculate \cite{HoroiNeacsu2016PSF} with good precision for most cases \cite{Stefanik2015,Kotila2012,StoicaMirea2013}. 
The "$\cdots $" sign represents other possible contributions, such as those of R-parity violating SUSY particle exchange \cite{Vergados2012,Horoi2013}, Kaluza-Klein modes \cite{Bhattacharyya2003,Deppisch2007,Horoi2013}, violation of Lorentz invariance, and equivalence principle \cite{Leung2000,Klapdor1999,Barenboim2002}, etc, that we neglected here. 
The $\eta^L_{N_R}$ term also exists in the seesaw type I mechanisms but its contribution is negligible if the heavy mass eigenstates are larger than 1 GeV \cite{Blennow2010}. We consider a seesaw type I dominance \cite{DevMitra2015} and we will neglect it here. 

For an easier read, we perform the following change of notation: $\left< \nu \right> = \left| \eta_{\nu} \right|$, $\left< \lambda \right>  = \left| \eta_{\lambda} \right|$ and $\left< \eta \right> = \left| \eta_{\eta} \right|\ $.

In this paper we provide an analysis of the two-electron relative energy and angular distributions for $^{124}$Sn, $^{130}$Te, and $^{136}$Xe using shell model NME that we calculate. 
The purpose of this analysis is to identify the relative contributions of $\eta_{\lambda}$ and $\eta_{\eta}$ terms in Eq. (\ref{flifetime}). 
A similar analysis for $^{82}$Se was done using QRPA NME in Ref. \cite{snemo2010} and with shell model NME in Ref. \cite{HoroiNeacsu2016prd}. 
As in Ref. \cite{HoroiNeacsu2016prd}, we start form the classic paper of Doi, Kotani and Tagasuki \cite{Doi1985} describing the neutrinoless double-beta decay process using a low-energy 
Hamiltonian that includes the effects of the right-handed currents. By simplifying some notations and ignoring the contribution from the $\eta^R_{N_R}$ term, which has the same energy and 
angular distribution as the $\eta_{\nu}$ term, the half-life expression \cite{Doi1985}  is written as
\begin{eqnarray}
\label{lifetime}
\nonumber \left[ T^{0\nu}_{1/2} \right] ^{-1}&=&\left|M_{GT}^{0\nu}\right| ^2 \left\{ C_{\nu^2} + C_{\nu\lambda}\textmd{cos}\phi_1 + C_{\nu\eta}\textmd{cos}\phi_2 \right.  \\
  &+& \left. C_{\lambda^2} + C_{\eta^2} +C_{\lambda\eta}\textmd{cos}(\phi_1 - \phi_2) \right\} ,
\end{eqnarray}
where $\phi_1$ and $\phi_2$ are the relative CP-violating phases (Eq. A7 of \cite{HoroiNeacsu2016prd}), and $M_{GT}^{0\nu}$ is the Gamow-Teller contribution of the light neutrino-exchange NME.
$C_{\alpha}$ are contributions from different mechanisms: $C_{\nu^2}$ are from the left-handed leptonic and currents, $C_{\lambda^2}$ from the right-handed leptonic and right-handed hadronic currents, and $C_{\eta^2}$ from the right-handed
 leptonic and left-handed hadronic currents.  $C_{\nu\lambda}$, $C_{\nu\eta}$ and $C_{\lambda\eta}$ contain the interference between these terms. These are defined as
\begin{eqnarray} \label{c-mechanism}
\nonumber C_{\nu^2}	&= C_1 \left< \nu\right> ^2, C_{\nu\lambda}	=& C_2\left< \nu\right>\left< \lambda\right>, 	C_{\nu\eta}	= C_3\left< \eta \right>\left< \nu\right>, \\
      C_{\lambda^2}	&= C_4\left< \lambda\right>^2, C_{\eta^2}	=& C_5\left< \eta \right>^2, 	\ \ \		C_{\lambda\eta}	= C_6\left< \eta \right>\left< \lambda\right> ,
\end{eqnarray}
where $C_{1-6}$ are combinations of nuclear matrix elements and phase-space factors (PSF).
Their expressions can be found in the Appendix B, Eqs. (B1) of Ref. \cite{HoroiNeacsu2016prd}. $M_{GT}^{0\nu}$ and the 
other nuclear matrix elements that appear in the expressions of the $C_i$ factors are presented in Eq. (B4) of Ref. \cite{HoroiNeacsu2016prd}. 

We write the differential decay rate of the $0^+\rightarrow0^+$ $0\nu\beta\beta$ transition as
\begin{equation}\label{diff-rate}
\frac{\textmd{d}^2W_{0^+\rightarrow0^+}^{0\nu}}{\textmd{d}\epsilon_1 \textmd{d}\text {cos}\theta_{12}} = \frac{a_{0\nu} \omega_{0\nu}(\epsilon_1)}{2\left( m_eR \right)^2} \left[A(\epsilon_1)+B(\epsilon_1)\text {cos}\theta_{12} \right]  .
\end{equation}
$\epsilon_1$ is the energy of one electron in units of $m_e c^2$, $R$ is the nuclear radius ($R=r_0 A^{1/3}$, with $r_0=1.2$fm), $\theta_{12}$ is the angle between the outgoing electrons, and the expressions for the constant $a_{0\nu}$ and the function $\omega_{0\nu}$ are 
given in the Appendix C, Eqs. (C2) and (C3) of Ref. \cite{HoroiNeacsu2016PSF}, respectively.
The functions $A(\epsilon)$ and $B(\epsilon)$ are defined as combinations of factors that include PSF and NME:
\begin{flushleft}
\begin{subequations}\label{ab}
\begin{flalign}
&A(\epsilon_1)\! = \! \left|N_1(\epsilon_1)\right|^2{\! + \!}\left|N_2(\epsilon_1)\right|^2{\! + \! }\left|N_3(\epsilon_1)\right|^2{\! + \!}\left|N_4(\epsilon_1)\right|^2 \! , \\
&B(\epsilon_1)\! = \! -2\text{Re}\left[ N_1^\star(\epsilon_1) N_2(\epsilon_1)\! + \! N_3^\star(\epsilon_1) N_4(\epsilon_1) \right].  
\end{flalign}
\end{subequations}
\end{flushleft}
 The detailed expressions of the $N_{1-4}(\epsilon_1)$ components are presented in Eqs. (B7) of Ref. \cite{HoroiNeacsu2016prd}.

We now express the half-life as follows
\begin{eqnarray} \label{half-life}
\nonumber \left[ T_{1/2}^{0\nu} \right]^{-1} &=& \frac{1}{\text{ln2}}\int \textmd{d}W_{0^+\rightarrow0^+}^{0\nu} =  \frac{a_{0\nu}}{\text{ln2} \left( m_eR \right)^2}\\
&\cdot& \int_1^{T+1} A(\epsilon_1) \omega_{0\nu}(\epsilon_1)\textmd{d}\epsilon_1,
\end{eqnarray} 
with the kinetic energy T defined as
\begin{equation}\label{t-energy}
 T=\frac{Q_{\beta\beta}}{m_e c^2}.
 \end{equation}

The integration of Eq. (\ref{diff-rate}) over $\epsilon_1$ provides the angular distribution of the electrons that we write as
\begin{eqnarray} \label{ec_angular}
 \nonumber \frac{\textmd{d}W_{0^+\rightarrow0^+}^{0\nu}}{\textmd{d}\Omega}&=&\frac{a_{0\nu}}{4\pi \left( m_eR \right)^2}\left[ 
 \int_1^{T+1} A(\epsilon_1)\omega_{0\nu}(\epsilon_1)\textmd{d}\epsilon_1 \right. \\ 
 &+& \left. \frac{\textmd{d}\Omega}{2\pi} \int_1^{T+1} B(\epsilon_1)\omega_{0\nu}(\epsilon_1)\textmd{d}\epsilon_1 \right],
\end{eqnarray}
where $d\Omega=2\pi d\text {cos}\theta_{12}$.
 
Integrating Eq. (\ref{diff-rate}) over cos$\theta_{12}$ provides the single electron spectrum. Similar to Ref. \cite{HoroiNeacsu2016prd}, we express the decay rate as a function of 
the difference in the energy of the two outgoing electrons, $\Delta t=(\epsilon_1 - \epsilon_2 )m_e c^2$, where
$ \epsilon_2=T+2-\epsilon_1$ is the kinetic energy of the second electron.
We can write the energy of one electron as
\begin{equation}
\epsilon_1=\frac{T+2+\frac{\Delta t}{m_e c^2}}{2}.
\end{equation}
Changing the variable, the energy distribution as a function of $\Delta t$ is
\begin{equation} \label{ec_energy}
 \frac{2\textmd{d}W_{0^+\rightarrow0^+}^{0\nu}}{\textmd{d}(\Delta t)}=\frac{2a_{0\nu}}{\left( m_eR \right)^2}
 \frac{\omega_{0\nu}(\Delta t)}{m_e c^2} A(\Delta t).
\end{equation}

\section{Results}
The formalism used in this paper is taken from Ref. \cite{HoroiNeacsu2016prd} where it was used to analyze the two-electron angular and energy distributions for $^{82}$Se, 
the baseline isotope of the SuperNEMO experiment \cite{snemo2010,nemo32014}. It was adapted from Ref. \cite{Doi1985} and Ref. \cite{SuhonenCivitarese1998} with some changes for 
simplicity, consistency, and updated with modern notations. Here we use it to analyze in detail the $0\nu\beta\beta$ decay two-electron angular and energy distributions 
for $^{124}$Sn, $^{130}$Te, and $^{136}$Xe. The nine NME required are calculated in this paper using the SVD shell model Hamiltonian 
\cite{Chong2012} in the $jj55$ model space that was thoroughly tested and validated for $^{124}$Sn in Ref. \cite{NeacsuHoroi2016}, and for $^{130}$Te and $^{136}$Xe in Ref. \cite{NeacsuHoroi2015}.
For an easier comparison to other results, we use a $g_A$ value of 1.254, we include shor-range correlations with CD-Bonn parametrization, finite nucleon size effects, and higher order corrections of the nucleon current \cite{HoroiStoica2010}. 
Should one change to the newer recommended $g_A$ value of 1.27 \cite{pdg-2014}, the NME results would change by only 0.5\% \cite{SenkovHoroiBrown2014} and the effective PSF 
(multiplied by $g_A^4$) by 5\%. This is negligible when compared to the uncertainties in the NME.

In Table \ref{tab-nme} we present the nine NME for $^{124}$Sn, $^{130}$Te and $^{136}$Xe calculated in this work using an optimal closure energy $\left< E \right> = 3.5$MeV that was 
obtained using a recently proposed method \cite{SenkovHoroi2014}. By using an optimal closure energy obtained for this Hamiltonian, we get $0\nu\beta\beta$ NME results in agreement with beyond closure approaches \cite{SenkovHoroi2013}.
 \begin{table}[htb] 
 \caption{The nine NME $^{124}$Sn, $^{130}$Te and $^{136}$Xe.}
 \begin{tabular}{@{}lccccccccc@{}} \hline \hline \label{tab-nme}
	  &$M_{GT}$&$M_{F}$&$M_{GT\omega}$&$M_{F\omega}$&$M_{GTq}$&$M_{Fq}$&$M_{T}$&$M_{R}$&$M_{P}$ \\ \hline 
$^{124}$Sn&1.85	& -0.47	& 2.05	&-0.46	& 1.79	& -0.27	& 0.07	&2.66	& 0.84   \\ 
$^{130}$Te&1.66	& -0.44	& 1.86	&-0.43	& 1.59	& -0.25	& 0.05	&2.56	& 0.95   \\ 
$^{136}$Xe&1.50	& -0.40	& 1.68	&-0.39	& 1.44	& -0.23	& 0.07	&2.34	& 0.94   \\ 
\hline
 \end{tabular}
 \end{table}

We calculate the integrated PSF that appear in the $C_\alpha$ components of  Eq. (\ref{lifetime}) using a new effective method \cite{HoroiNeacsu2016PSF} in agreement with the latest results and that was tested for 11 nuclei. 
The largest difference from Ref. \cite{Stefanik2015} for our three isotopes of interest is of about 16\% for $G_8$ for $^{136}$Xe. One should keep in mind that the expressions for the two-electron angular and energy 
distributions contain energy dependent (un-integrated) PSF, and not the integrated PSF that are found in tables. 
Should one use the formalism of Ref. \cite{Doi1985}, differences of about 88\% are expected in the case of $G_8$ for $^{136}$Xe.
The values for the nine integrated PSF are presented in Table \ref{tab-psfs}. The results shown include the $g_A^4=1.254$ constant, such that
$G_1$=$G_{01}^{0\nu}g_A^4$ in Eq. (\ref{flifetime}) and $G_{[1,9]}$=$G_{[01,09]}g_A^4$ of Ref. \cite{Stefanik2015}.

\begin{table}[htb] 
 \caption{The nine PSF expressed in $\left[ yr^{-1}\right] $}
 \begin{tabular}{@{}l@{}c@{}c@{}c@{}c@{}c@{}c@{}c@{}c@{}c@{}} \hline \hline \label{tab-psfs}
		&$G_1$		&$G_2$		&$G_3$		&$G_4$		&$G_5$		&$G_6		$&$G_7$		&$G_8$		&$G_9$ \\ 
		&$\cdot 10^{14}$&$\cdot 10^{14}$&$\cdot 10^{14}$&$\cdot 10^{15}$&$\cdot 10^{13}$&$\cdot 10^{12}$&$\cdot 10^{10}$&$\cdot 10^{11}$&$\cdot 10^{11}$ \\ \hline 
$^{124}$Sn	&\ 1.977	&\ 4.184	&\ 1.248	&\ 3.909	&\ 6.685	&\ 3.648	&\ 2.749	&\ 2.585	&\ 0.731	\\
$^{130}$Te	&\ 3.122	&\ 8.026	&\ 2.092	&\ 6.267	&\ 10.18	&\ 5.335	&\ 4.340	&\ 4.261	&\ 1.106	\\
$^{136}$Xe	&\ 3.188	&\ 7.798	&\ 2.114	&\ 6.372	&\ 10.79	&\ 5.464	&\ 4.458	&\ 4.522	&\ 1.099	\\ \hline
 \end{tabular}
\end{table}

The $C_{i}$ factors ($i=1, \ldots , 6$) of Eq. (\ref{c-mechanism}), representing combinations of NME and PSF, are presented in Table \ref{tab-ci}. As one can clearly see, the $C_5$ term that 
appears in the $\eta$ mechanism is the largest. This is because of the $G_7$, $G_8$, and $G_9$ PSF displayed in Table \ref{tab-psfs}.

\begin{table}[htb]
 \caption{The $C_{i}$ factors ($i=1, \ldots , 6$) corresponding to Eq.  (\ref{c-mechanism}) expressed in $\left[ yr^{-1}\right] $. }
  \begin{tabular}{lcccccc} \hline \hline  \label{tab-ci}
		&$C_1$		&$C_2$		&$C_3$		&$C_4$		&$C_5$		&$C_6$ 			\\ 
		&$\cdot 10^{14}$&$\cdot 10^{14}$&$\cdot 10^{12}$&$\cdot 10^{13}$&$\cdot 10^{9}$	&$\cdot 10^{13}$	\\ \hline	 
$^{124}$Sn   	&2.67		&-1.27		&5.74		&0.54		&1.34		&-0.71	\\ 
$^{130}$Te   	&4.25		&-4.17		&8.95		&1.64		&2.26		&-2.34	\\ 
$^{136}$Xe   	&4.36		&-2.22		&9.15		&1.04		&2.24		&-1.34	\\ 
\hline
\end{tabular} 
\end{table}

To test the possibility of disentangling the right-handed contributions in the framework of the left-right symmetric model, we consider three theoretical cases:
that of the mass mechanism denoted with $\nu$ and presented with the black color in the figures, the case of $\lambda$ mechanism dominance in competition with $\nu$, denoted with $\lambda$ 
and displayed with the blue color, while the last case is that of the $\eta$ mechanism dominance in competition with $\nu$, denoted with $\eta$ and displayed with the red color.
This color choice is consistent throughout all the figures. 

Considering the latest experimental limits \cite{Barry2013,Stefanik2015} from the $^{76}Ge \ 0\nu\beta\beta$ half-life, we select a value for the mass mechanism parameter $\nu$ that corresponds
to a light neutrino mass of about 1 meV, while the values for the $\lambda$ and $\eta$ effective parameters are chosen to barely dominate over the mass mechanism. Should their values be 
reduced four times, their contributions would not be distinguishable from the mass mechanism.

\begin{table}[htb]
 \caption{The neutrino parameter values chosen for the $0\nu\beta\beta$ mechanisms described in the text.}
 \begin{tabular}{@{}lccc@{}} \hline \hline \label{cases}
				&$\left< \nu\right>$	&$\left< \lambda\right>$& $\left< \eta\right>$\\ \hline
mass mechanism ($\nu$)		&$2\cdot 10^{-7}$	&$0$			&$0$			\\
lambda mechanism ($\lambda$)	&$2\cdot 10^{-7}$	&$2\cdot 10^{-7}$	&$0$			\\
eta mechanism ($\eta$)		&$2\cdot 10^{-7}$	&$0$			&$2\cdot 10^{-9}$	\\\hline
\end{tabular}
\end{table}

We consider four combinations for the CP phases $\phi_1$ and $\phi_2$ (each one being 0 or $\pi$) that can influence the half-lives and the two-electron distributions. 
The maximum difference arising from the interference of these phases produces the uncertainties that are displayed as bands in the figures, changing the amplitudes and the shapes. 
As the mass mechanism does not depend on $\phi_1$ and $\phi_2$, there is no interference, and it is represented by a single thick black line. 
Because the mass mechanism is the most studied case in the literature, one may consider it as the reference case.

\begin{figure}[htb]
\includegraphics[width=0.99\linewidth]{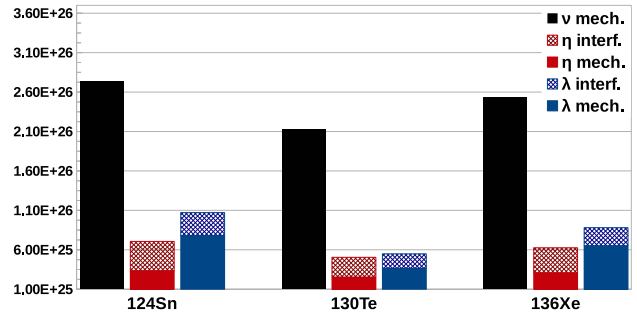}
\caption{The calculated lifetimes and their uncertainties (the hatched bars) form the interference of the unknown CP phases.}
\label{fig_lifetimes} 
\end{figure}

The calculated half-lives of $^{124}$Sn, $^{130}$Te and $^{136}$Xe are presented in Table \ref{half-lives}. Their values can be obtained either from Eq. (\ref{lifetime}), or from Eq. (\ref{half-life}).
The maximum differences from the interference phases produces the intervals.
For an easier comparison of the half-lives and the uncertainties, we also plot them in Fig. \ref{fig_lifetimes}.  
One can notice that the inclusion of the $\lambda$ or $\eta$ contributions reduces the half-lives.
\begin{table}[htb] 
 \caption{Calculated half-lives $(T_{1/2})$ intervals for each mechanism expressed in years. The range of the interval corresponds
 to the uncertainty in the CP phases $\phi_1$ and $\phi_2$ in Eq. (\ref{lifetime}).}
 \begin{tabular}{cccc} \hline \hline \label{half-lives} 
	  &$\nu$		&$\lambda$			&$\eta$	 \\ \hline 
$^{124}$Sn&$2.73\cdot 10^{26}$	&$[0.78,1.07]\cdot 10^{26}$	&$[3.34,7.06]\cdot 10^{25}$	\\ 
$^{130}$Te&$2.12\cdot 10^{26}$	&$[3.64,5.49]\cdot 10^{25}$	&$[2.52,5.05]\cdot 10^{25}$	\\ 
$^{136}$Xe&$2.53\cdot 10^{26}$	&$[6.50,8.80]\cdot 10^{25}$	&$[3.07,6.25]\cdot 10^{25}$	\\ 
\hline
 \end{tabular}
\end{table}

\begin{figure}[htb]
\includegraphics[width=0.99\linewidth]{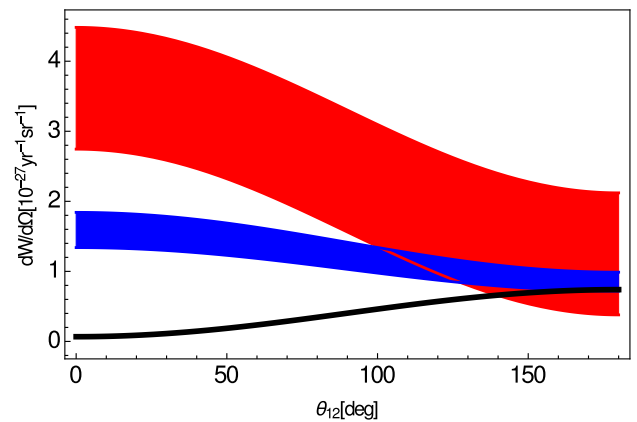}
\caption{ Electrons angular distribution for $^{124}$Sn.}
\label{124sn_ang} 
\end{figure}

\begin{figure}[htb]
\includegraphics[width=0.99\linewidth]{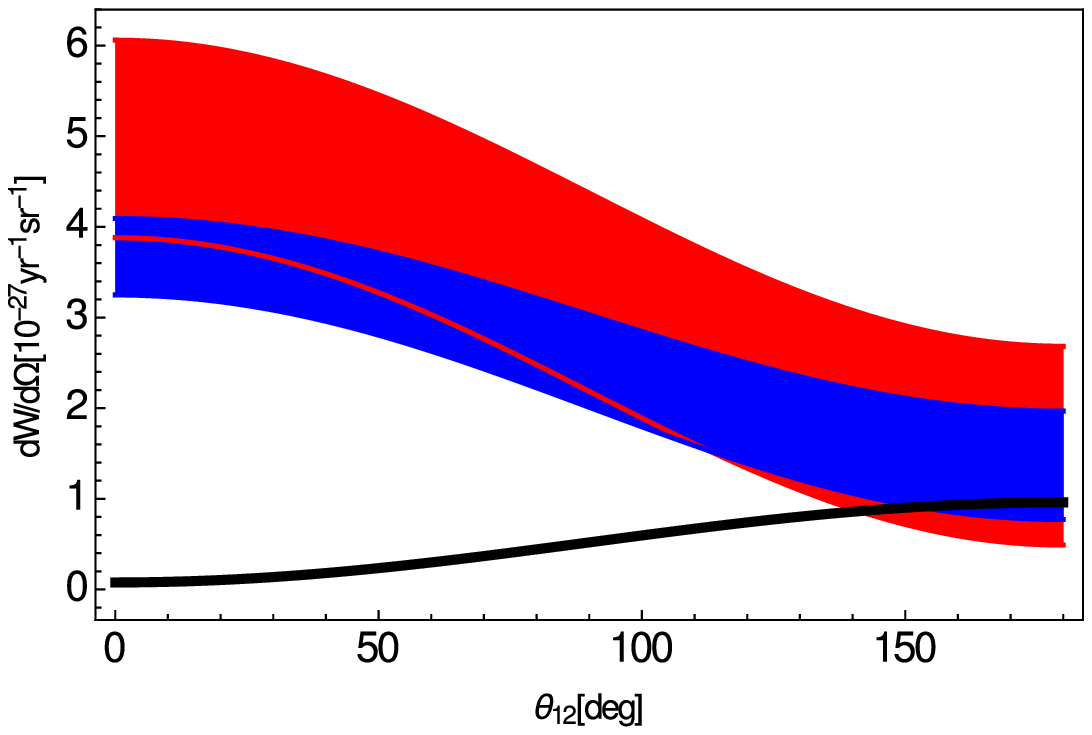}
\caption{ Electrons angular distribution for $^{130}$Te.}
\label{130te_ang} 
\end{figure}

\begin{figure}[htb]
\includegraphics[width=0.99\linewidth]{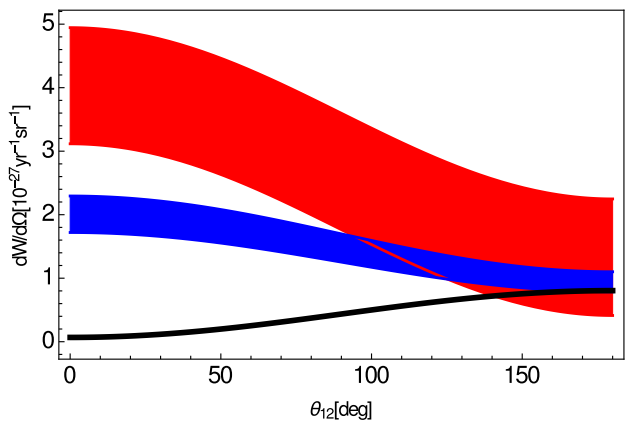}
\caption{ Electrons angular distribution for $^{136}$Xe.}
\label{136xe_ang} 
\end{figure}

The shapes of the two-electron angular distributions of Eq. (\ref{ec_angular}) could be used to distinguish between the mass mechanism and the $\lambda$ or the $\eta$ mechanisms. However, many recorded 
events (tens or more) are needed for a reliable evaluation, and even then one can face difficulties due to the unknown CP phases.
The $^{124}$Sn angular distribution is presented in Fig. \ref{124sn_ang}. One can see that $\lambda$ (blue bands) and $\eta$ (red bands) exhibit similar shapes, differing in amplitude, and 
opposite to that of the mass mechanism (black line). In the case of $^{130}$Te, the same is to be expected, but the $\lambda$ and $\eta$ bands overlap due to the unknown phases, as seen in 
Fig. \ref{130te_ang}. The $^{136}$Xe angular distribution is very similar to that of $^{124}$Sn and is presented in Fig. \ref{136xe_ang}.

\begin{figure}[htb]
\includegraphics[width=0.98\linewidth]{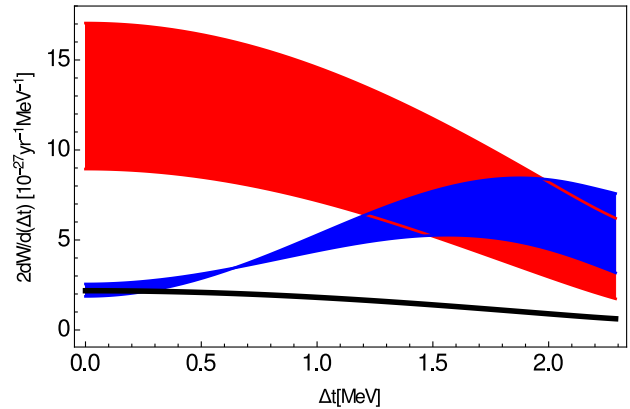}
\caption{ Electrons energy distribution for $^{124}$Sn.}
\label{124sn_en} 
\end{figure}

\begin{figure}[htb]
\includegraphics[width=0.98\linewidth]{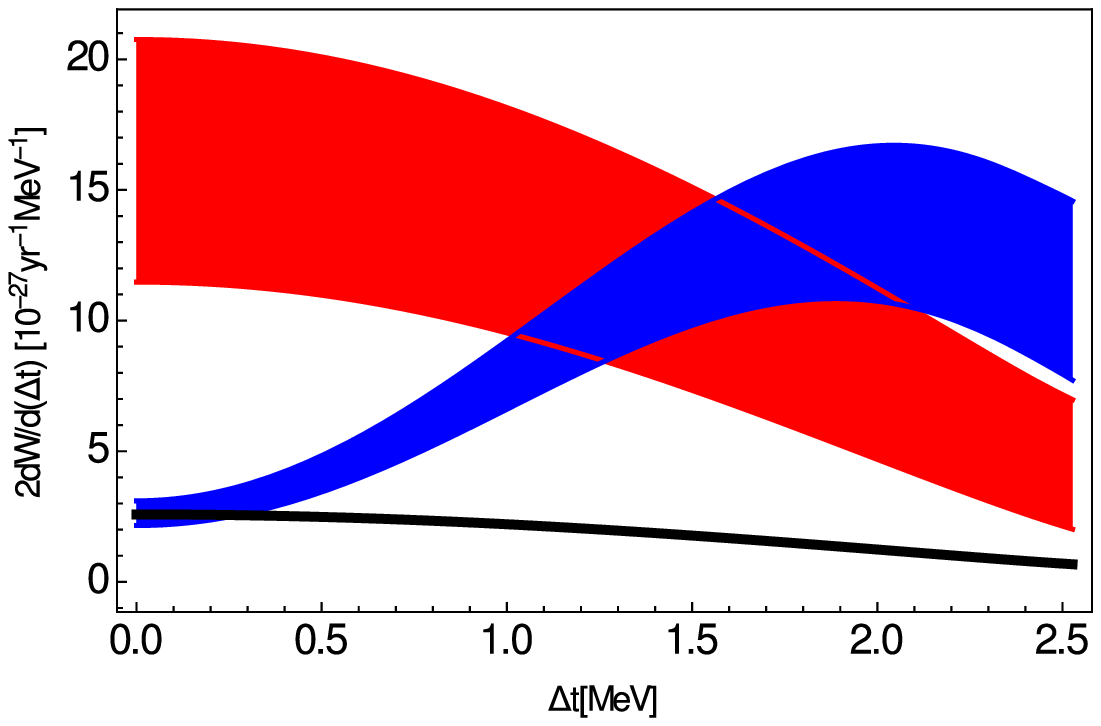}
\caption{ Electrons energy distribution for $^{130}$Te.}
\label{130te_en} 
\end{figure}

In principle, the $\lambda$ and the $\eta$ contributions could be identified in the shapes of the two-electron energy distributions. While the tails of the distributions (when the difference 
between the energy of one electron and that of the other is maximal) overlap, the starting points (when both electrons have almost equal energies) are very different for the $\lambda$ from the $\eta$
mechanism. Fig. \ref{124sn_en} shows the energy distribution for $^{124}$Sn. The $^{130}$Te energy distribution is presented in Fig. \ref{130te_en}. For $^{136}$Xe we find an energy 
distribution very similar to that of $^{124}$Sn, like in the case of the angular distributions.

\begin{figure}[htb]
\includegraphics[width=0.98\linewidth]{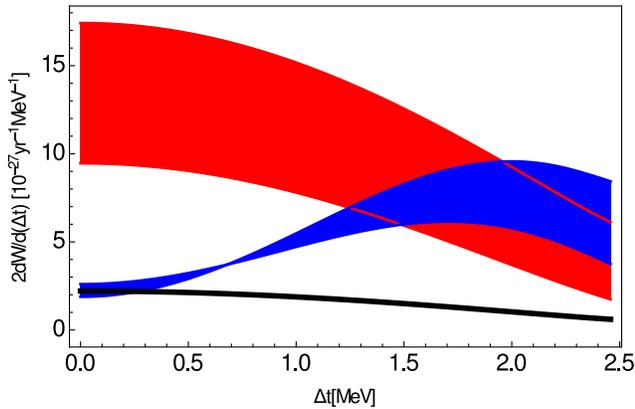}
\caption{ Electrons energy distribution for $^{136}$Xe.}
\label{136xe_en} 
\end{figure}

\begin{figure}[ht!]
{\includegraphics[width=0.98\linewidth]{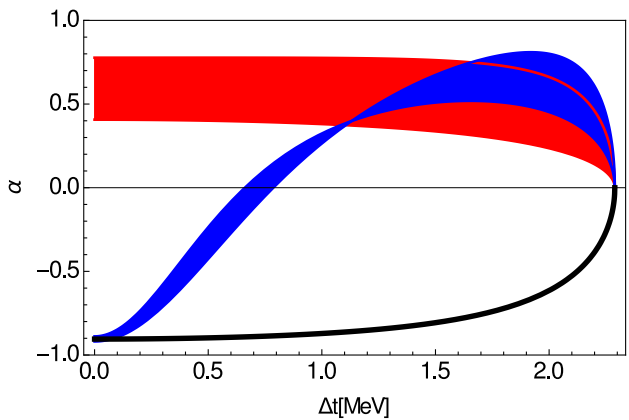}}
\caption{The angular correlation coefficient for $^{124}$Sn.}
\label{124sn_alpha}
\end{figure}

\begin{figure}[ht!]
{\includegraphics[width=0.98\linewidth]{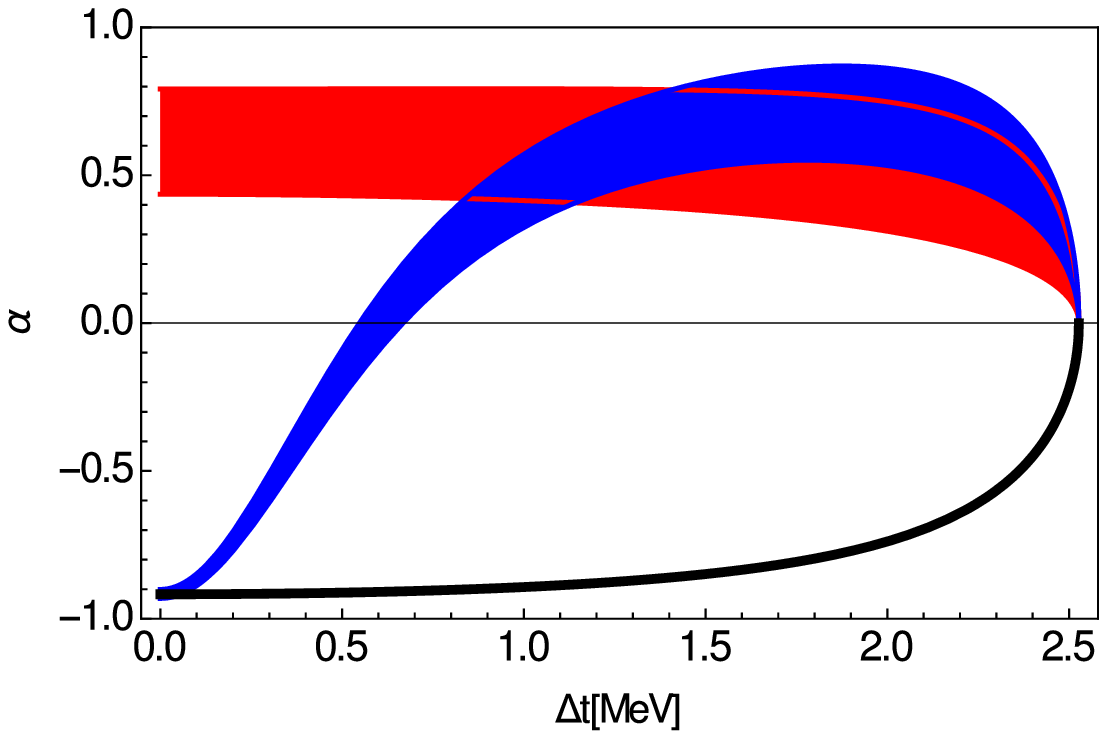}}
\caption{The angular correlation coefficient for $^{130}$Te.}
\label{130te_alpha}
\end{figure}

\begin{figure}[ht!]
{\includegraphics[width=0.98\linewidth]{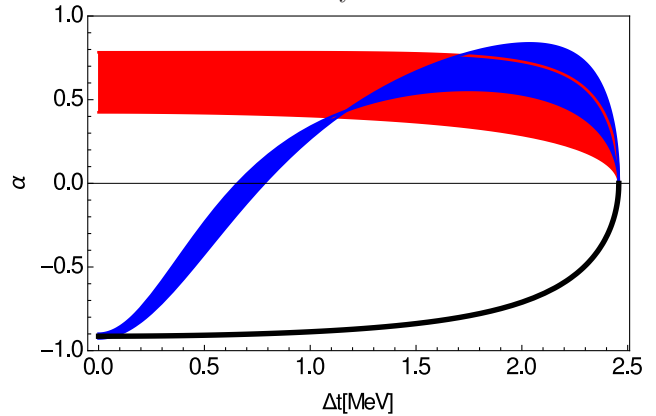}}
\caption{The angular correlation coefficient for $^{136}$Xe.}
\label{136xe_alpha}
\end{figure}

To further aid with the disentanglement of the $\lambda$ and $\eta$ mechanisms, we provide plots of the angular correlation coefficient, $\alpha = B(\epsilon)/A(\epsilon)$ in our Eq. (\ref{diff-rate}).
This may help reduce the uncertainties induced by the unknown CP phases (see e.g. Figs. 6.5 -  6.9 of \cite{Doi1985} and Fig. 7 of \cite{Stefanik2015}). 
From $\alpha(\Delta t)$, one may also obtain a clearer separation from the mass mechanism over a wide range of energies. The angular correlation coefficient for $^{124}$Sn is presented in Fig. \ref{124sn_alpha}.
The same behavior can be identified in Fig. \ref{130te_alpha} for $^{130}$Te and in Fig. \ref{136xe_alpha} for $^{136}$Xe. 

\section{Conclusions}

In this paper we report shell model calculations necessary to disentangle the mixed right-handed/left-handed currents contributions 
(commonly referred to as $\eta$ and $\lambda$ mechanisms) from the mass mechanism in the left-right symmetric model. We perform an analysis of these
contributions by considering three theoretical scenarios, one for the mass mechanism, another for the $\lambda$ dominance in competition with the mass mechanism, and 
a scenario where the $\eta$ mechanism dominates in competition with the mass mechanism. 

The figures presented support the conclusions \cite{Doi1985,HoroiNeacsu2016prd} 
that one can distinguish the $\lambda$ or $\eta$ dominance over the mass mechanism from the shape of the two-electron angular distribution, while one can 
discriminate the $\lambda$ from the $\eta$ mechanism using the shape of the energy distribution and that of the angular coefficient. 
The tables and the figures presented also show the uncertainties related to the effects of interference from the unknown CP-violating phases.

We show our results for phase space factors, nuclear matrix elements and lifetimes for $0\nu\beta\beta$ transitions of $^{124}$Sn, $^{130}$Te, and $^{136}$Xe to ground states.
In the case of the mass mechanism nuclear matrix elements, we obtain results which are consistent with previous calculations \cite{NeacsuHoroi2015,HoroiNeacsu2016prd}, where the 
same SVD Hamiltonian was used. Similar to the case of $^{82}$Se \cite{HoroiNeacsu2016prd}, the inclusion of the $\eta$ and $\lambda$ mechanisms contributions tends to decrease the half-lives.

The phase-space factors included in the analysis of life-times and two-electron distributions are calculated using a recently proposed accurate effective method \cite{NeacsuHoroi2016} that provides results 
very close to those of Ref. \cite{Stefanik2015}. Ref. \cite{Stefanik2015} takes into account consistently the effects of the realistic finite size proton distribution in the daughter nucleus, but it does not provide all the energy-dependent phase-space contributions necessary for our analysis. 

Consistent with the calculations and conclusions we obtained for $^{82}$Se \cite{HoroiNeacsu2016prd}, if the $\eta$ mechanism exists, it may be favored to compete with the mass mechanisms because of the larger contribution from the phase space factors. 

Finally, we conclude that in experiments where outgoing electrons can be tracked, this analysis is possible if enough data is collected, generally of the order 
of a few tens of events. This may be beyond the realistic capabilities of the current experiments, but should a positive neutrinoless double-beta decay measurement be achieved,
it is expected that more resources could be allocated to improve the statistics and the variety of investigated isotopes.

\begin{acknowledgments}
Support from the NUCLEI SciDAC Collaboration under U.S. Department of Energy Grant No. DE-SC0008529 is acknowledged.
MH also acknowledges U.S. NSF Grant No. PHY-1404442 and U.S. Department of Energy Grant No. DE-SC0015376.
\end{acknowledgments}

\bibliographystyle{apsrev}
\bibliography{new_ahep}

\end{document}